# Ship Detection in SAR Images with Human-in-the-Loop

Hecheng Jia, *Member, IEEE*, Feng Xu, *Senior Member, IEEE*

*Abstract*—Synthetic aperture radar (SAR) has been extensively utilized in maritime domains due to its all-weather, all-day monitoring capabilities, particularly exhibiting significant value in ship detection. In recent years, deep learning methods have increasingly been utilized for refined ship detection. However, learning-based methods exhibit poor generalization when confronted with new scenarios and data, necessitating expert intervention for continuous annotation. Currently, the degree of automation in human-machine collaboration within this field, especially in annotating new data, is not high, leading to labor- and computation-intensive model iteration and updates. Addressing these issues, a ship detection framework in SAR images with human-in-the-loop (HitL) is proposed. Incorporating the concept of HitL, tailored active learning strategies are designed for SAR ship detection tasks to present valuable samples to users, and an interactive human-machine interface (HMI) is established to efficiently collect user feedback. Consequently, user input is utilized in each interaction round to enhance model performance. Employing the proposed framework, an annotated ship database of SAR images is constructed, and the iteration experiments conducted during the construction demonstrates the efficiency of the method, providing new perspectives and approaches for research in this domain.

## I. Introduction

AS the field of earth observation and satellite imaging technology advances, remote sensing images are characterized by significantly wider and higher resolution, offering broader coverage and richer target details. Different from optical sensors, Synthetic Aperture Radar (SAR) possesses unique all-weather and all-time capabilities, playing an important role in various fields such as agriculture, military, and urban planning. In recent years, focusing on common remote sensing targets like airplanes, ships, and buildings, there has been an emergence of a substantial body of research on SAR image-based target detection. Among these, SAR ship detection has become indispensable in fisheries management, port planning, and maritime surveillance.

Traditional Non-DL methods primarily utilize constant false alarm rate (CFAR) and its variants [1]-[4]. These methods detect ship targets based on the intensity difference between targets and clutter, offering advantages such as high processing speed and no requirement for data learning. However, they are limited in complex scenarios, such as coastal areas. With the widespread application of deep learning in computer vision, an increasing number of researches [5]-[9] have introduced deep learning into ship detection task in SAR images, achieving significant improvements in effectiveness.

Contemporary mainstream deep learning methods rely heavily on extensive data and annotations, creating models through one-time learning for specific tasks, and perform well in tasks where data distribution is similar to the training samples. However, in practical applications, SAR images are received as a data stream. When encountering new tasks, it's challenging to ensure that the trained models exhibit good generalization on unknown data. Therefore, algorithms need the capability to progressively acquire, update, accumulate, and utilize new data. In response to this, some researches [10]-[13] use continual learning methods to enhance the adaptability and scalability of models to new scenarios and data. Despite these efforts, the involvement of annotation personnel is unavoidable.

Human-machine collaboration is a method that leverages human-computer interaction to enhance the efficiency of learning from new data. It utilizes the strengths of both humans and machines to achieve more efficient, accurate, and reliable data processing and decision-making. Remote sensing is one of the important application scenarios for human-machine collaboration. As new observational data continuously emerges, the challenge lies in how to process and identify new scenes and types of targets. However, the level of automation in human-machine collaboration in remote sensing is currently not high, particularly in the area of new data annotation. Inefficient data annotation directly impacts the adaptation speed and robustness of deep learning-based models to new scenes and categories. On the one hand, traditional manual annotation can be costly and unreliable, especially for SAR images, which require extensive expert experience for accurate annotation. On the other hand, existing automated or semi-automated approaches lack effective utilization of human feedback and guidance, making it difficult to ensure annotation quality and consistency. Therefore, designing effective human-machine interaction mechanisms to achieve collaborative annotation is an urgent problem to be addressed.

In the general field of machine learning, to enhance the

This work was supported in part by the National Natural Science Foundation of China.

The authors are with the key lab for information science of electromagnetic waves (MoE), Fudan University (e-mail: fengxu@fudan.edu.cn).

automation of human-machine collaboration, some researches introduce the concept of "human-in-the-loop (HitL)" [14], [15], integrating human feedback into the modeling process and strengthening human-machine interaction in machine learning, achieving certain results. For example, Fan et al. [16] propose an intelligent annotation method that combines active learning and visual interaction, using iterative user annotation to detect network anomalies. Liu et al. [17] develop a human-machine interactive model based on reinforcement learning for pedestrian re-identification tasks. Wallace et al. [18] introduce a human-machine adversarial generation framework that improves the model's understanding through human guidance. These studies have explored the application of the HitL method in human-machine collaboration, achieving effects in their respective tasks. However, there are significant differences in the practical application of this method to specific tasks. For instance, when applied to ship target detection in SAR images, it's necessary to consider challenges such as large image range, sparse targets, and high false alarm rates. Designing targeted active learning strategies and efficient human-machine interface (HMI) is key to applying HitL approach to specific tasks.

Addressing the aforementioned challenges, a HitL-based detection framework is proposed for SAR images, enhancing the efficiency of human-machine collaboration in ship detection tasks through iterative training and human-machine interaction. On one hand, a tailored active learning approach is utilized to select valuable samples for the user, reducing the workload of annotating new data and improving annotation quality. On the other hand, an interactive human-machine interface (HMI) is designed for user in the loop to efficiently correct detection results which guide the iterative optimization of the model and creating better training and evaluation data. The innovation and effectiveness of the work are primarily reflected in the following aspects:

1) An active learning strategy is designed suitable for ship detection tasks in SAR images, enhancing the efficiency of user annotation and correction of samples, and assisting users in completing tasks more quickly.

2) An efficient interactive HMI is constructed, using manual feedback to guide the iterative optimization of the model, enabling the model to achieve the highest accuracy in ship detection task through the learning loop.

3) A SAR ship detection database of SAR images is built through the proposed loop framework, including high-quality ship location annotations and a variety of fine-grained ship types. The iterative experiments conducted during the construction process demonstrate the efficiency of the proposed method.

The following content of the paper will first provide a detailed introduction to the designed loop framework in Section II. Sections III and IV then respectively discuss the core components within the loop: the selection of valuable samples and the design of the interactive HMI. Finally, Section V describes the application of the proposed framework in the construction of the SAR ship database.

## II. THE PROPOSED FRAMEWORK

The proposed HitL framework integrates expert experience into the machine learning process, guiding the optimization of the model by correcting current model results. This approach differs from the traditional one-time learning and the autonomous learning. Fig. 1 illustrates the differences in the workflow among these approaches.

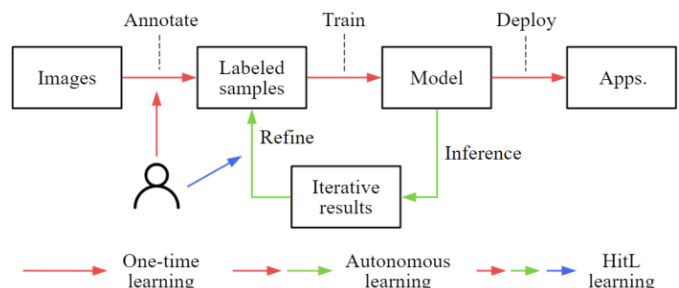

Fig. 1. Comparison of three learning frameworks.

1) One-Time Machine Learning. As shown by the red arrows in Fig. 1, this process includes data collection, annotation, model training, and application. Human involvement is limited to initial data annotation, providing labeled data for supervised training. The machine's contribution to humans lies in the prediction results during application. This learning process does not form a closed loop.

2) Autonomous Learning. Illustrated by the red and green arrows, this differs from one-time learning in that the inference results of new tasks are incorporated into the next round of model training, forming a closed loop in machine learning. In this loop, the annotation and training of new tasks are done through self-supervised or unsupervised methods, with human involvement only in the initial pre-training samples.

3) HitL Framework. The learning process of this framework, as shown by the red, green, and blue arrows, involves humans in the iterative loop of machine learning. Users fine-tune the results of each iteration and feed their expertise back into the algorithm model, forming an optimization process based on human feedback.

The detail structure of the designed HitL framework is illustrated in Fig. 2, with the specific process steps as follows:

Step 1: At the onset of the task application, partial existing data and a pre-trained model are utilized for transfer learning, resulting in the initial detection model.

Step 2: The trained model is employed to complete the current ship detection task, with the predicted results presented to expert users.

Step 3: An active learning strategy is applied to select valuable samples and feedback them to the users for verification and correction.

Step 4: Users correct the sample predictions through the designed HMI, and the results are stored in the sample database.

Step 5: Training samples are chosen based on the attributes of samples in the sample database, and the model is updated and applied to new tasks in the task flow.

Step 6: Steps 2-5 are repeated.

The iterative update design of the loop continuously enhances the model's accuracy, each iteration encompassing three key processes: inference, refinement, and training, corresponding to step 2, steps 3 and 4, and step 5, respectively.

**Inference.** As depicted in Fig. 2(b), this module consists of several sub-modules. To address the issues of extensive scenes and sparse targets in SAR images, the preprocessing module includes sea-land segmentation and CFAR preprocessing. This step filters out a significant portion of land and clutter areas, thus improving inference efficiency. For the challenge of high false alarm rates in SAR ship detection, the subsequent parts include a detection network and a false alarm discrimination network, both powered by deep learning-based methods. The results are then visualized for user review and refinement.

**Refinement.** Targeting both ship detection and false alarm discrimination, as shown in Fig. 2(c), ship detection refinement involves adjusting the ship bounding boxes, whereas false alarm refinement entails modifying the subcategories of false alarms. The refined targets are then stored in the sample database.

**Interactive Training.** Incorporating a false alarm discrimination module, the training process, as illustrated in Fig. 2(d), involves both the detection and discrimination networks. Iterative training samples are derived from user-refined data and a selection of historical samples.

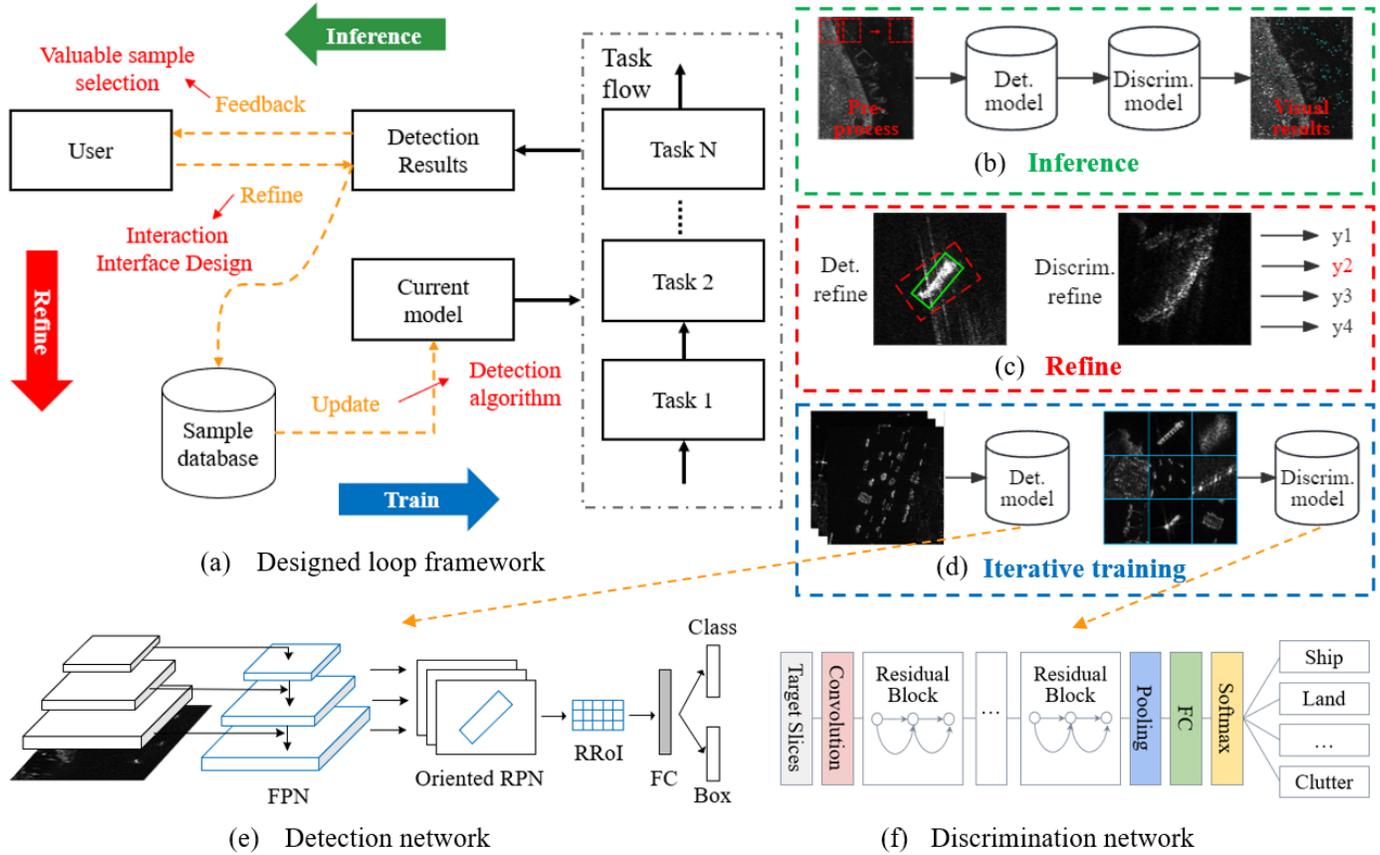

Fig. 2. The proposed framework for ship detection in SAR images.

Human-machine interaction is the foundation of human-in-the-loop machine learning, as shown in Fig. 2. Around this interaction, the designed learning loop encompasses three main aspects: detection algorithm, valuable sample selection, and HMI design. The detection algorithm is involved in the training and inference processes, as previously described, and is divided into detection network and discrimination network.

The detection network is implemented based on the oriented detection structure of Oriented R-CNN [19], as shown in Fig. 2 (e). It includes a Feature Pyramid Network (FPN) [20] module, Oriented Region Proposal Network (RPN), rotated Region of Interest (RoI) extraction, and classification and regression sub-networks. The training is based on fixed-size scene slices containing ships, while the inference is based on pre-processed candidate ship areas.

The false alarm discrimination network is realized based on the ResNet [21], as illustrated in Fig. 2 (f). The input is target scale slices that, after feature extraction, employ Softmax to output normalized classification confidence. The classification categories include ships as positive samples and fine-grained false alarms as negative samples.

Building upon the algorithms used in the loop, the subsequent two sections will separately discuss valuable sample selection and the design of the HMI.

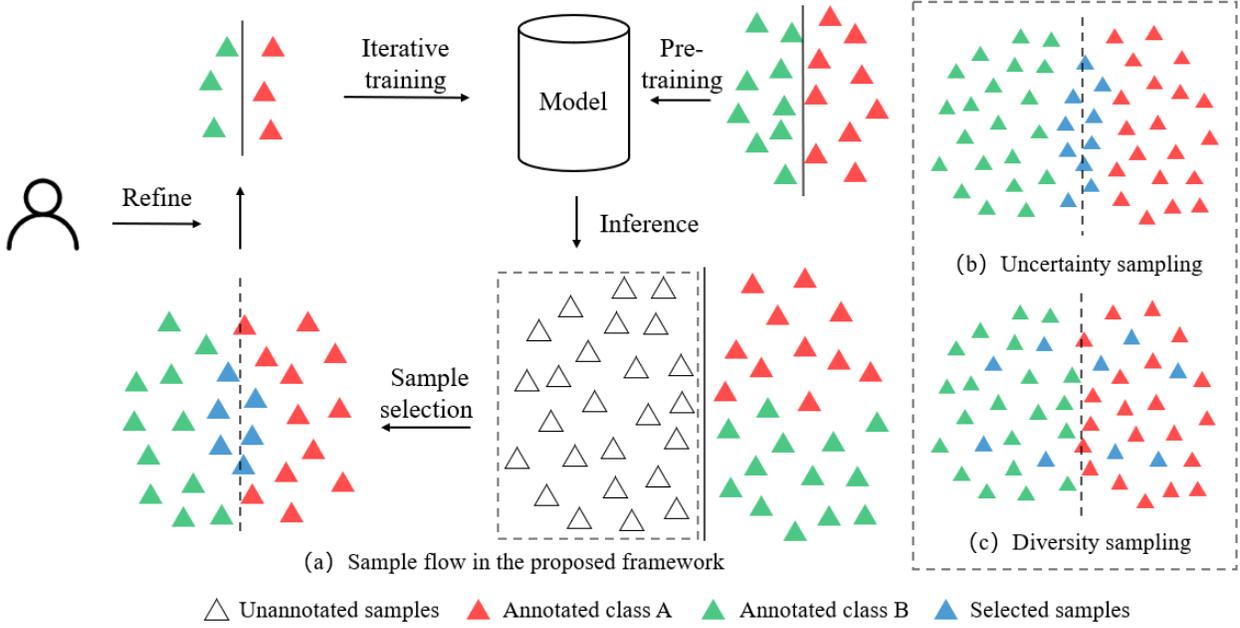

Fig. 3. Sample flow and sampling strategy in the proposed framework

## III. VALUABLE SAMPLE SELECTION

In practical applications, from the perspective of annotation status, annotated samples, unannotated samples and selected samples constitute the sample database for designed learning loop and form the data foundation for ship detection task. The utilization and flow of samples in the loop are illustrated in Fig. 3(a). Among these, the selected samples, which are recommended to users for fine-tuning in each iteration, are referred to as valuable samples in the loop. The selection process of these valuable samples involves filtering the most valuable detection results to be corrected by the user, thereby reducing the annotation workload.

The selection of valuable samples should satisfy criteria of uncertainty and diversity. As illustrated in the right of Fig. 3, green and red represent two categories of samples, with blue indicating the samples recommended for annotation. In the figure, (b) represents the selection of uncertain samples, which are typically the most unstable predictions from the previous model iteration, and (c) depicts random sample selection, ensuring the diversity of training data.

In this paper, an active learning approach is employed to filter the results of algorithm inference, which includes sampling based on result uncertainty, result diversity, and automatic contrast sampling.

### A. Uncertainty Sampling

The algorithm within the loop encompasses two types of deep learning models: detection and discrimination, and confidence is the fundamental metric for assessing the uncertainty of prediction results. Different sampling strategies are designed for each type of models, as detailed below.

*1) Discrimination Task*

The discrimination task accomplishes the function of classification, which includes two scenarios in the proposed loop: the classification head of the detection network and the discrimination network. The classification head is used for binary classification of foreground and background, while the discrimination network is utilized for ship and fine-grained false alarm classification. Both networks employ the last fully connected layer to produce the logits for each category and use the Softmax to normalize the result probability distribution, defined as follows:

$$S_j = \frac{e^{z_j}}{\sum_{k=1}^{K} e^{z_k}} \quad (1)$$

Here, $S_j$ represents the probability score of the $j$-th class, $z_j$ is the original logit for the $j$-th class, and $K$ is the total number of categories.

After processing with the activation function, each image sample will have multiple confidence outputs, corresponding to the number of categories, and the background makes the total count one more than the number of categories. Softmax is sensitive to the base number used, and directly using the highest confidence score might not adequately represent the influence of the remaining categories on prediction uncertainty. To mitigate this sensitivity, margin of confidence sampling is used to assess the uncertainty of the results. The uncertainty of discrimination for a single sample is defined as follows:

$$U_r = 1 - (S_{\text{top1}} - S_{\text{top2}}) \quad (2)$$

Here, $U_r$ is the uncertainty score, with $S_{\text{top1}}$ and $S_{\text{top2}}$ being the highest and second-highest confidence scores predicted for all categories in a single sample. The difference between these two scores is taken and subtracted from 1 to normalize, resulting in the final uncertainty for the sample.

*2) Detection Task*

In addition to the classification subnet, the head of detection network also includes a box subnet, which predicts the position of ship targets. For the box subnet, since it employs a regression model that does not automatically output confidence values. However, the accuracy of bounding box predictions is also an important reference for measuring the uncertainty of detection results. Consequently, the proposed method designs an uncertainty sampling method for predicted bounding boxes, utilizing multiple prediction outputs and the intersection over union (IoU) calculation. To enhance the efficiency of the learning loop, a single model is used to output multiple inference results, generating them using a dropout strategy.

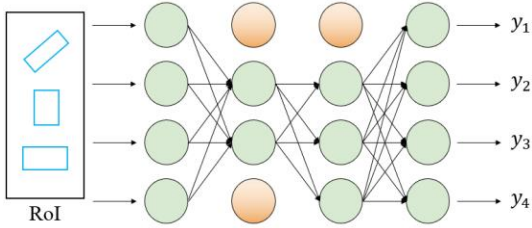

Fig. 4. Dropout strategy of the bounding box regression subnetwork.

As shown in Fig. 4, the dropout is typically used in the training process of convolutional neural networks, where it randomly disables some network nodes to prevent overfitting. In the proposed method, dropout is similarly applied during the inference stage. By randomly disabling neurons in the bounding box regression subnet during each inference, multiple predictions are generated, and the IoU among these results can reflect the uncertainty of sample localization.

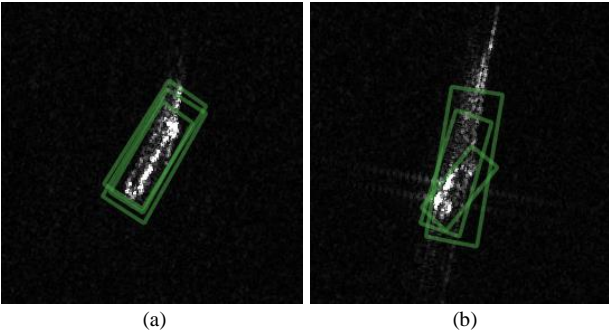

(a) (b)

Fig. 5. Illustration of Generating multiple predicted bounding boxes. (a) High confidence target. (b) Low confidence target.

Fig. 5 illustrates multiple inference results for two targets in a SAR ship detection task. It is evident that the variability of the predicted bounding boxes in (b) is greater than in (a), indicating that refining the detection results in (b) holds more value for iterative training.

To quantify uncertainty, multiple results for the same target are used to calculate the mean IoU (mIoU), which is then utilized to determine the uncertainty of the detected box:

$$C_i = \frac{1}{n-1} \sum_{j=1, j\neq i}^{n} \text{IoU}(B_i, B_j) \quad (3)$$

$$U_d = 1 - (C_{top1} - C_{top2}) \quad (4)$$

In this formula, $C_i$ is the confidence score of the $i$-th bounding box, $n$ is the number of inferences, which is the number of bounding boxes contained in each result, $B_i$ is the current bounding box, and $U_d$ is the uncertainty of the current target's predicted bounding box. $C_{top1}$ and $C_{top2}$ represent the bounding boxes with the highest and second-highest confidence scores, respectively. Thus, each ship will have multiple confidence scores for predicted bounding box. If a prediction for a target does not yield a result in a certain instance, the confidence score is set to 0. Then, using the margin of confidence sampling, the uncertainty of each ship's location can be determined. Additionally, if a sample is selected for refinement, the bounding box with the highest confidence is returned.

Upon determining the uncertainties for detection and discrimination, the final uncertainty, $U_t$, is calculated using a harmonic mean approach similar to the F-score:

$$U_t = \frac{(1+\beta^2) \cdot U_r \cdot U_d}{\beta^2 \cdot U_r + U_d} \quad (5)$$

In this formula, $\beta$ represents a weighting factor, which is set to 1 in this context.

After assigning uncertainty scores to all results, the samples are ranked. In each iteration, the top 10% of samples are selected for feedback to the user. Additionally, to provide a larger pool of valuable samples, an oversampling of the detection results is conducted. This is achieved by adjusting the confidence threshold to be 0.1 less than the optimal threshold.

*B. Diversity Sampling*

Unlike uncertainty sampling, which focuses on samples with the greatest variation, diversity sampling aims to prevent bias in sample selection. Considering the complexity and efficiency of SAR ship detection, the proposed method employs real-scenario diversity sampling.

Real scenarios in SAR ocean imagery exhibit two characteristics: large scene and uneven sample distribution. To address the issue of large scene, in addition to sampling target areas, false alarms are also sampled to increase diversity, given the high false alarm rate in SAR images. Then, the unevenness of samples in the loop algorithm is characterized by both category imbalance and uneven ship distribution. The sampling methods are as follows.

*1) Category Imbalance*

For ship detection task in proposed loop, the term "category imbalance" refers to the ship category and multiple fine-grained false alarm subcategories. A category-balanced selection method is introduced to increase the diversity of classes in the samples recommended. This method selects samples based on the weight of each category. Given a dataset containing $n$ different categories, with each category $i$ having $x_i$ samples, the total number of samples and the maximum number of samples in any category are first calculated. Then, a value $\mu$ is computed to serve as the basis for the weights, defined as follows:

$$\mu = \frac{max(x_1, x_2, \ldots, x_n)}{\sum_{i=1}^{n} x_i} \quad (6)$$

Subsequently, based on $\mu$, the number of samples $x_i$ in each category, and the total number of samples, the weight $w_i$ for each category $i$ is calculated as follows:

$$w_i = max\left(1, ln\left(\mu \cdot \frac{\sum_{j=1}^n x_j}{x_i}\right)\right) \quad (7)$$

Here, $w_i$ aims to increase the weight of minority categories while decreasing the weight of majority categories, thereby mitigating the impact of data imbalance. A logarithmic function is used to smooth the weighting factors. Subsequently, the number of samples to be selected for each category can be calculated, creating a more balanced set of valuable samples. The number of samples $s_i$ to be selected for each category is calculated as follows:

$$s_i = min(\text{p} \cdot w_i \cdot x_i, x_i) \quad (8)$$

In this formula, p is a predetermined proportion representing the desired sample proportion to be selected from each category, which is set to 10% in this case. The number of samples to be selected for each category is the smaller value between $\text{p} \cdot w_i \cdot x_i$ and $x_i$, ensuring that the number of selected samples does not exceed the total number of samples in that category.

*2) Uneven Distribution of Ships*

Due to the characteristics of large-scale scenes, the uneven distribution of ship positions is particularly pronounced. A sliding window approach is employed to select training scenes for the detection network, retaining only those windowed scenes that contain ships, and achieving a more balanced distribution of positive and negative samples. However, within the sliding windows, there is still an issue of uneven sample distribution. Fig. 6 illustrates scenes within sliding windows where the number of samples is uneven. Some scenes show clustered arrangement of targets, while others contain only a single target.

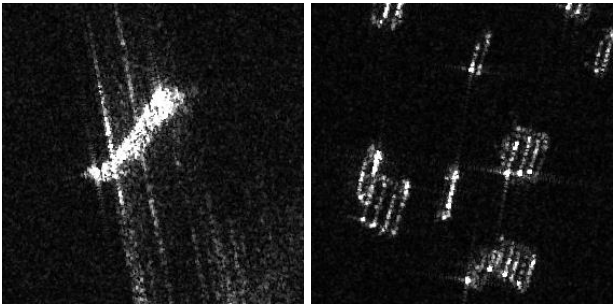

Fig. 6. Illustration of uneven sample quantity in sliding windows.

To ensure diversity in the sample distribution of detection tasks in real-world scenarios, the proposed method employs balanced sampling based on the number of samples within each sliding window. Fig. 7 shows the distribution statistics of detection results in sliding window slices for a SAR ship detection task during a certain iteration. The blue and yellow bars represent the original number of slices, and it is evident that most slices contain fewer than five samples.

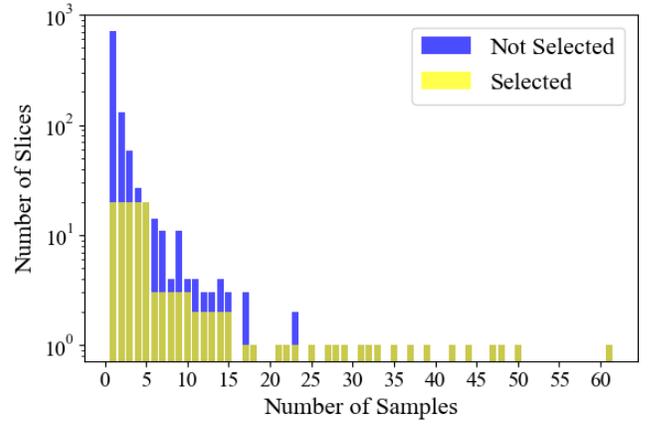

Fig. 7. Statistics of sample quantity in sliding windows.

The proposed method groups window slices into intervals of five samples, counts the number of slices in each group, and then uses the balancing method described in "Category Imbalance" to calculate the number of slices to be sampled from each interval. Subsequently, random sampling is performed within each group. The effects of this sampling, shown as yellow bars in Fig. 7, indicate that the method increases the sampling rate of slices from scenes with fewer samples, thus enhancing the overall diversity of the training data.

*C. Automatic Contrast Sampling*

In practical applications, given the typically large size of SAR images, it is challenging for users to cover all detection results in each correction iteration. To enhance the efficiency of user annotations and model iterations, multiple rounds of iteration processes are performed on some large-sized images within the designed learning loop. For scenarios involving multiple predictions of the same data across iterations, starting from the second iteration, an automatic contrast sampling is designed to select valuable samples.

TABLE I
MEANING OF SAMPLE TYPES IN CONTRASTIVE SAMPLING

| Type | Implication | Equal to RGT |
|---|---|---|
| TP | Correctly detected ships | √ |
| FP | Incorrectly detected ships or falsely identified false alarms | |
| TN | Correctly identified false alarms | √ |
| FN | Missed detections of ships | |

Automatic contrast refers to automatic correction based on results from the previous iteration. The automatic correction strategy is determined by evaluating the current results with the results corrected in the previous round, where the previous round's corrected results are used as the referenced ground truth (RGT). The evaluation of results considers both discrimination accuracy and bounding box prediction accuracy, where predictions with correct categories and box IoU greater than 0.5 with the RGT are considered positive samples. After evaluation, four types of samples are identified: True Positive (TP), False Positive (FP), True Negative (TN), and False Negative (FN). Table I illustrates the meaning of the four types, based on the assumption that the results corrected in the previous round are considered the truth.

Among these, FP and FN are inconsistent with the RGT and represent the valuable samples that require particular attention from the user. To guide the user's focus more effectively towards these targets, different color markings are used: TP and TN are indicated with intensified colors, while FP and FN are marked with subdued colors. Fig. 8 shows the visualization results provided to the user. Compared to the RGT, light green and light red represent correctly identified ships and negative samples with a lower probability of error; dark blue indicates missed detections or category errors, and dark red marks false alarms with a higher error probability.

Similarly, to increase recall and reduce missed detections, the predicted results are oversampled, retaining those with a confidence level reduced by 0.1 from the optimal confidence.

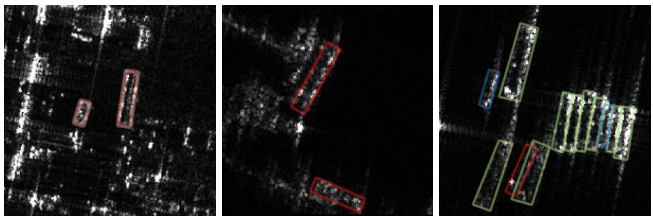

Fig. 8. User attention guidance based on contrastive sampling.

## IV. INTERACTIVE HUMAN-MACHINE INTERFACE

In each iteration of the proposed learning loop, users perform sample verification and correction through the interactive HMI. The interface is fundamental for user involvement in machine learning and needs to consider both operational design and user experience. This section outlines the design of HMI tailored for SAR images, facilitating user corrections of detection results. It comprises two main aspects: effective operational design and sample attribute design. The user interface is built upon the open-source software of QGIS [22].

### A. Interaction Operation Design

For ship detection, the visualization results provided to users include the targets' categories and their locations, represented as polygons with labels. Based on the representations, four common operations are designed: Add, Modify, Delete, and Browse, and the usage scenarios are illustrated in Table II.

TABLE II
OPERATION DESIGN AND USAGE SCENARIOS

| Operation | Usage Scenarios |
| --- | --- |
| Add | Mark missed ships during the review process |
| Modify | Verify and correct the labels or bounding boxes of ships, including marking negative samples |
| Delete | Remove difficult samples |
| Browse | Examine specific areas of the entire scene |

Unlike generic annotation system, the operations designed within the interface cater to both positive and negative samples. In current mainstream deep learning-based detectors, the sampling of negative samples during training is typically random. Guided and valuable negative samples can effectively enhance the algorithm's capability to discriminate false alarms, especially in large-sized SAR images where ships are relatively sparse. In the loop, the negative samples primarily originate from discrimination model within the algorithm.

The correction operations are designed to be tolerant of user input, not imposing strict limitations on the number or type of corrections. However, users still need to follow certain correction principles during the operation process. These principles can enable the designed learning loop to train high-accuracy models more quickly , as outlined below:

1) Verification and Annotation of Missed Targets Around Corrected Samples: The window slices where the corrected sample is located are included in the next training round. As shown in Fig. 9(a), unmarked positive samples in the scene are treated as negative samples, affecting the training of the model.

2) Elimination of Redundant Results in Multiple Predictions for a Single Target: As depicted in Fig. 9(b), the designed loop system may produce multiple detection boxes for a single target, mainly due to the IoU threshold and category confusion. Users need to retain a single result.

3) Fine-Grained Negative Sample Annotation: During the correction process for negative samples, users should annotate specific category for negative samples as much as possible. This information can be used for better fine-grained false alarm filtering, as shown in Fig. 9(c), where false alarm results are marked with specific coastal categories.

4) Removal of Uncertain Difficult Samples: Fig. 9(d) shows an example of a difficult sample that is hard to judge, highlighted in yellow. Difficult samples can affect sample quality and should be removed from the iterative training process.

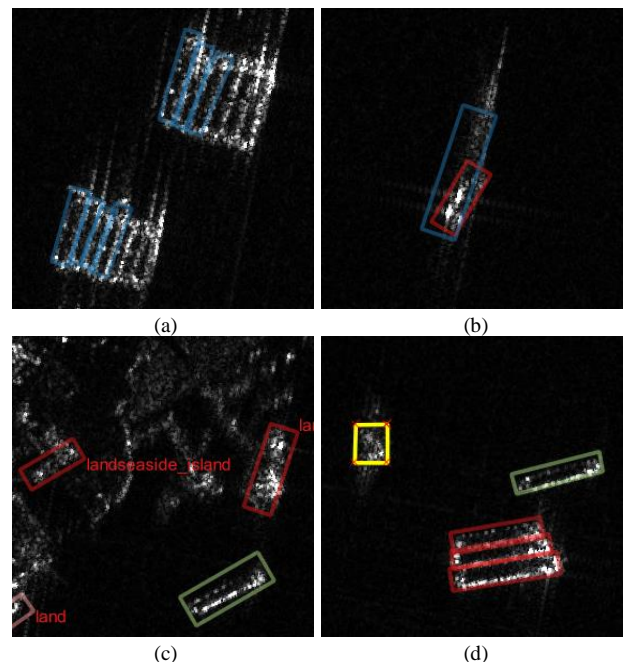

Fig. 9. Illustration of situations corresponding to refine principles.

### B. Sample Attribute Design

For ship detection task, samples contain two basic pieces of information: category and location. In the proposed learning loop, it is necessary to record user operations and sample states, hence an expanded attribute design for samples has been developed and stored in the sample database. The designed sample attributes, as shown in Table III, include the following: creation and modification times to record whether the user has corrected the predicted results; category attributes divided into positive and negative samples, each with corresponding confidences; uncertainty scores and diversity indicators to highlight the importance of sample correction and verification; the source field to indicate whether the result is from inference or automatic correction; and the status field as a reference for the sample's state during user operations, including uncorrected, automatically corrected, and manually corrected.

TABLE III
DESCRIPTION OF SAMPLE ATTRIBUTES

| Field | Definitions | Field | Definitions |
| --- | --- | --- | --- |
| name | Detection Category | source | Creation Source |
| score | Detection Confidence | cat_scores | Discrimination Confidence |
| uid | Unique Identifier | cat_name | Discrimination Category |
| create_time | Creation Time | status | Current Status |
| edit_time | Modification Time | uncertainty | Uncertainty Score |
| positive | Positive/Negative Sample Indicator | diversity | Diversity Indicator |

The attribute information designed is recorded in text format in attribute files corresponding to each image and stored in JSON format. As each iteration process generates prediction results, multiple attribute files with time series are produced for each image. The proposed method uses timestamps to uniquely identify each attribute file of an image, with the final attribute file serving as the ground truth annotation file.

Images and their attribute files are stored to form the sample database of the designed loop. Each training round selects training samples based on the attribute files corrected in the current image, choosing both positive and negative samples corrected by users and the results of diversity sampling as the foundational data for training. Additionally, to prevent model underfitting and catastrophic forgetting during incremental learning, t a sample replay strategy is employed, which involves adding some historical images and their corresponding attribute files from the database to the training set.

## V. EXPERIMENTS AND DATABASE CONSTRUCTION

The proposed method is applied to the task of constructing a ship database of SAR images, enabling practicality verification while creating a high-quality database that includes precise ship location annotations and extensive category information. This section first details the iteration experiments conducted during the database construction process. It then describes the method of automatically assigning ship categories using Automatic Identification System (AIS) information, and finally, a comprehensive overview of the constructed database is provided.

### A. Iteration Experiments

The database is constructed based on 30 GF-3 scenes covering maritime scenarios. Within the proposed learning loop, users only verified and corrected the ship location and false alarm samples. The fine-grained categorization of the ships will be addressed in the next subsection B.

#### 1) Data Config

The study divides 30 images into three batches, each containing 10 scenes. Batches 1 and 2 serve for training, while Batch 3 is designated for testing and evaluation. Batches 1 and 2 each undergo three rounds of result refinement, totaling six rounds of refinement. After each refinement round, the results are used to retrain the model and assess its performance. The evaluation is conducted on a fixed set of 10 test scenes, with metrics including the number of samples refined, refinement percentage, recall, and false alarm rate, employing a confidence threshold of 0.8 for assessment. To reduce workload in practice, the experiment utilizes the SSDD [23] dataset for pre-training the model, which is then applied for initial inference in Batch 1.

#### 2) Algorithm Config

The algorithm in the loop includes the ship detection network and false alarm discrimination network mentioned in Section B. The inference process of the algorithm incorporates a preprocessing module, consisting of sea-land segmentation and CFAR operations, to enhance detection efficiency. The configuration of the preprocessing module remains unchanged during the iterative process. The detection network uses ResNet50 as backbone, and each iteration involves 36 training epochs. The scene sliding window size for training and inference is 1024x1024 pixels. The false alarm network also employs ResNet50 for feature extraction, undergoing 200 training epochs per iteration. The network's predictions include seven categories: ship, land, coastal, bridge, sea surface, clutter, and the other. Both training and inference of the model in the loop are performed using a single NVIDIA RTX3090.

#### 3) Human-Machine Interface Config

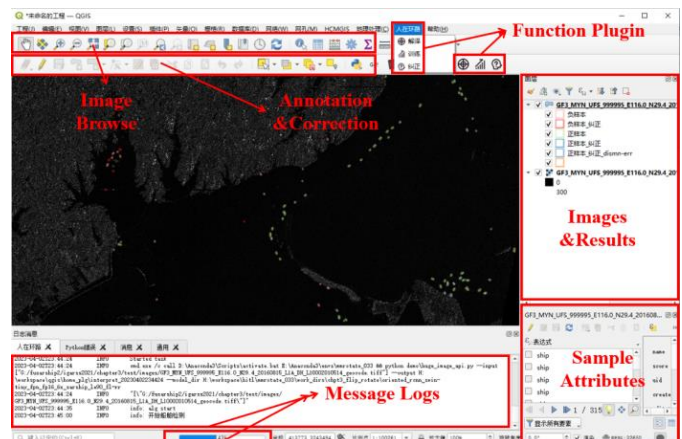

Fig. 10. Human-machine interaction interface based on QGIS.

The experiment utilizes QGIS software as the basic HMI system. A loop function plugin, developed based on the design from Section IV, is integrated into the QGIS software, as shown in Fig. 10. The training of the algorithm model is conducted offline, while inference and fine-tuning are accomplished using the plugin's features. In each round within the loop, three annotators are arranged to refine the results. Sample uncertainty and diversity serve as references, with each round of refinement involving about 10% of samples, without strict limitations.

Additionally, to enhance annotation efficiency and data quality, the interaction system incorporates two auxiliary features: optical map assistance and AIS information assistance. As illustrated in Fig. 11(a) and (b), the optical map is presented using open-source maps loaded through QGIS, while AIS information is converted to a QGIS-compatible format and displayed in the corresponding layer.

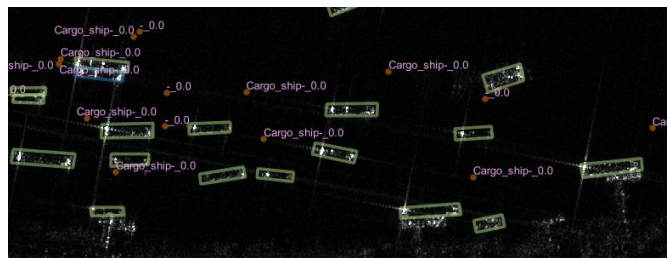

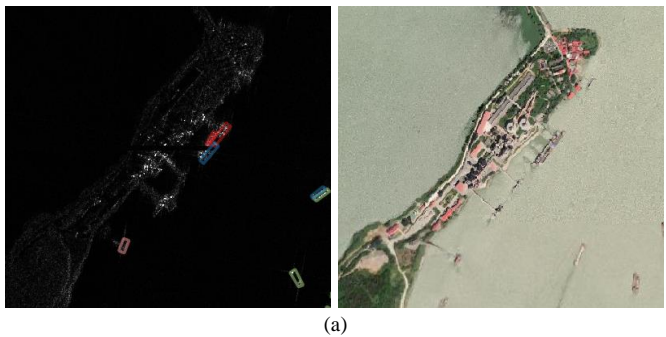

Fig. 11. Auxiliary annotation function. (a) Optical map assistance. (b) AIS information assistance.

*4) Experiment Results*

The experimental results are summarized in Table IV. The "Batch" column indicates the refinement batch, for example, "1-2" refers to the second refinement of Batch 1. Batch 0 represents the scenario and results with the pre-trained model, which already had ground truth annotations, so no sample refinement was performed. Batch 3 was used for the fixed test set, hence no evaluation of recall and false alarm rate is conducted and only overall refinement results are tallied. "Training Count" refers to the number of samples used for iterative training after the end of the current refinement, and it includes both positive and negative samples. "Recall" and "False Alarm Rate" are the results for the fixed test set (Batch 3).

TABLE IV
ITERATION EXPERIMENT RESULTS

| Batch | Refinement Count | Refinement Proportion (%) | Training Set | Training Count | Recall (%) | False Alarm Rate (%) |
|---|---|---|---|---|---|---|
| 0 | - | - | Pre-training Data | 2456 | 59.59 | 28.06 |
| 1-1 | 121 | 9.38 | Batch 1 | 1290 | 65.62 | 23.62 |
| 1-2 | 108 | 7.87 | | 1372 | 70.21 | 19.84 |
| 1-3 | 66 | 4.71 | | 1401 | 72.99 | 17.91 |
| 2-1 | 168 | 10.21 | Batches 1 and 2 | 3047 | 81.37 | 13.16 |
| 2-2 | 145 | 8.04 | | 3204 | 86.13 | 12.26 |
| 2-3 | 96 | 5.10 | | 3285 | 89.44 | 11.10 |
| 3 | 235 | 17.70 | Batches 1, 2 and 3 | 4613 | - | - |

The results show that the performance in Batch 0 was relatively poor. This is due to the significant difference between the pre-training data and the GF-3 data used in the experiment. Therefore, pre-training data is only used in the first iteration and not included in subsequent training.

**Analysis of Annotation Workload.** For multiple refinements of the same batch, the refinement percentage of first round is around 10%. As iterations progressed, the model's learning capacity for the same batch gradually optimized, leading to a further decrease in the proportion of user refinements, with only about 5% in the third round. Additionally, after incorporating the new data of Batch 2, the percentage of refinement increases. This is due to the model's initial lack of effective learning for new scenes and data. However, as iterative training progresses, this proportion gradually decreases, as evidenced by the results of the second and third iterations in Batch 2. The overall annotation workload for Batches 1 and 2 is around 20%, and for Batch 3, it is 17%, lower due to the enhanced performance of the model trained on the first two batches. Overall, the method significantly reduces the user's annotation workload.

**Analysis of Algorithm Performance.** As shown in the evaluation results of the test set in Table IV, with the increased number of refinements, the Recall gradually improved, and the false alarm rate decreased. This improvement is mainly due to continuous correction of samples and the addition of new batch data. For the same batch, each round of refinement gradually has a smaller impact on model performance, with the first round showing the greatest improvement. This is partly because the proportion of user refinements is highest in the first round and partly because multiple rounds of training on the same batch data has saturated the model's performance on that batch.

Additionally, the first round of refinement after adding a new batch significantly enhances model performance, as shown by the first round of metric results for Batch 2. The addition of new scenes and samples is crucial for improving model performance, and the proposed method progressively enhances the model's generalization capability in streaming tasks.

Moreover, the overall false alarm rate for this batch of images is not high, partly because preprocessing operations eliminate most land and sea clutter areas. As the number of refinements increases, the accumulation of false alarm discrimination task samples grows, further reducing the model's false alarm rate.

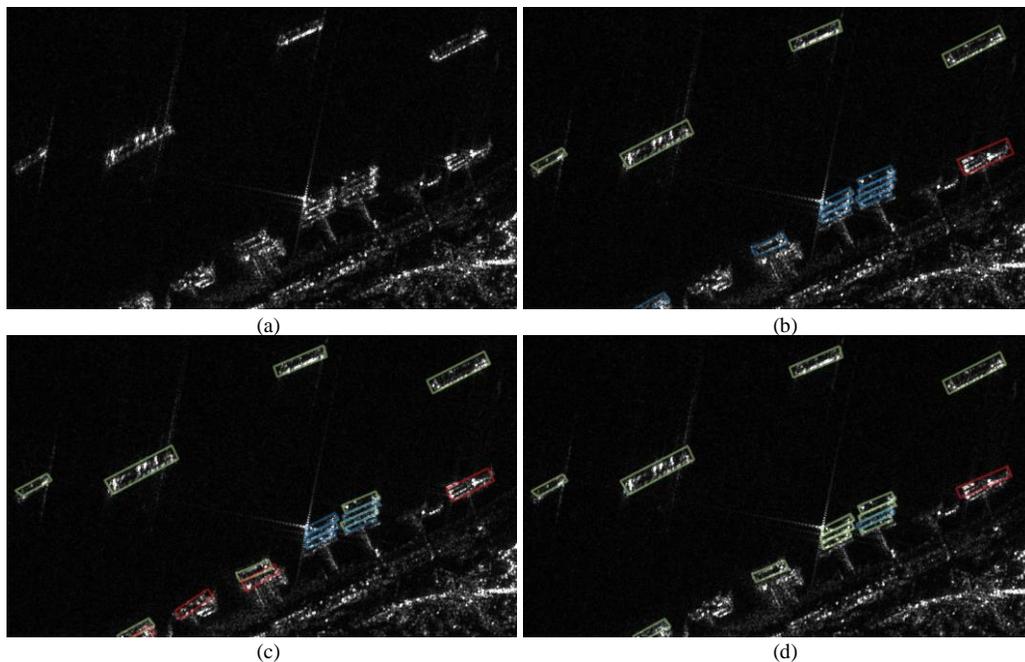

Fig. 12. Visualization of iterative results in coastal areas. (a) represent the original image. (b), (c) and (d) represent the results of Batch 0, Batch 1-3, and Batch 2-3, respectively.

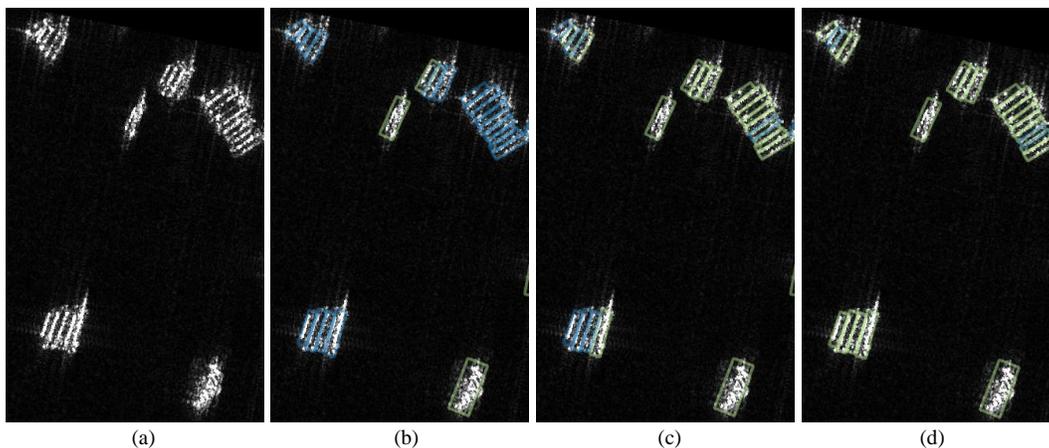

Fig. 13. Visualization of iterative results in densely arranged areas. (a) represent the original image. (b), (c) and (d) represent the results of Batch 0, Batch 1-3, and Batch 2-3, respectively.

This section also presents visualization results from different iterative rounds. Fig. 12 shows a coastal scene, while Fig. 13 displays a densely arranged ship area. In these figures, green indicates correct detections, blue represents missed detections, and red denotes false alarms. As shown in both figures (b), the pre-trained model exhibits average performance with many missed detections, especially in densely packed areas. However, it is capable of detecting a certain number of ships, reducing the initial annotation workload. After 6 rounds of iterative refinement through Batches 1 and 2, the model performance significantly improves in both coastal and offshore scenes, as illustrated in (c) and (d) of the two figures.

For SAR ship detection in complex nearshore scenarios, the algorithm tends to generate numerous false alarms. The loop's targeted addition of negative sample categories for refinement gradually reduces coastal false alarms as iterations progress. In offshore densely packed ship areas, as users progressively annotate missed detections, the number of undetected ships steadily decreases.

*B. Annotation of Fine-Grained Ship Categories*

After completing the annotation of ship positions using the proposed loop method, this paper employs AIS information to automatically assign fine-grained categories to ships based on their locations. AIS is an automatic tracking system for ships, exchanging digital data with ships and coastal stations, including important information such as ship category, precise location, heading, and speed.

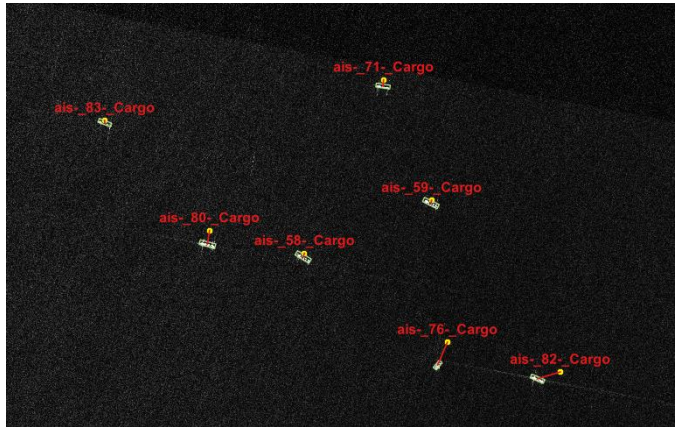

Fig. 14. Illustration of AIS-Ship Detection Matching.

This study first acquires an AIS dataset covering the time and location of the captured images and preprocessed it to filter unique information for each ship within the range and time of each image. Then, for each individual ship, its position and time are determined by interpolating the closest two records from all its received records. Subsequently, the Hungarian algorithm, commonly used in target detection and tracking fields, is employed. Here, AIS information is treated as tracked objects, matching AIS ship coordinate points with the center points of all detection boxes.

Fig. 14 shows the visualization of the matching results in QGIS, with yellow points representing AIS ship coordinates, green boxes as ship detection boxes, and red lines indicating the matching relationships. It is evident that the employed matching method accurately assigns category information to the ships.

The 30 images in the study contain a total of 3783 positive sample ships. Out of these, AIS successfully matched 2014 targets. The missing targets are primarily due to incomplete AIS coverage for certain image areas and times, as well as some ships not transmitting AIS information. For the categories that are matched, secondary matching is performed by scraping information from public ship databases. The unmatched 1769 targets are labeled as the 'Unknown' category. The matched 2014 targets encompass 8 major categories and 47 subcategories, as show in Fig. 15. The 8 major categories include Cargo, Fishing, Tanker, Passenger, Tug, Dredger, Law Enforcement, and the Other. The category distribution shows a clear long-tail pattern, with Cargo being the most prevalent, followed by Fishing and Tanker, and the other categories being less numerous. 'Other' consists of types with fewer than 10 instances. Moreover, this finer categorization provides a foundation for researching specific SAR ship features and models.

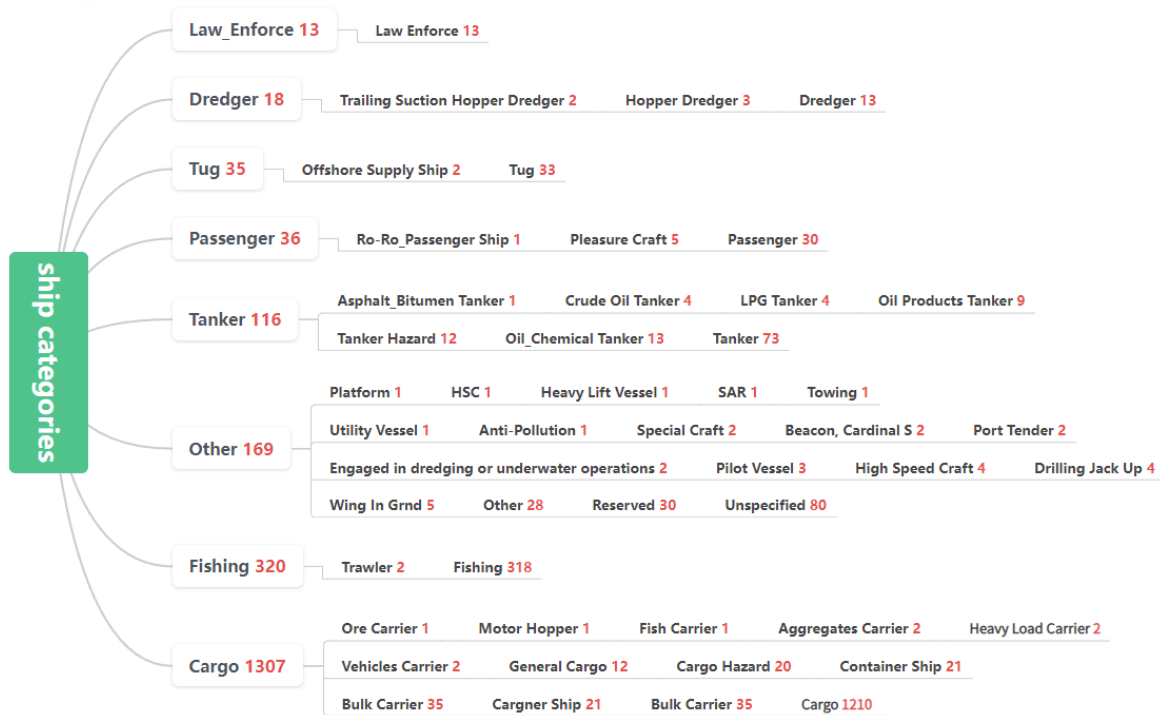

Fig. 15. Distribution of fine-grained ship categories.

C. Statistics of Constructed Database

Utilizing the proposed loop method and AIS category matching, a ship detection database of GF-3 SAR images is constructed. The data encompasses 30 ultra-fine strip mode images from GF-3 satellite, with a resolution of 1 meter. The locations include areas with heavy ship traffic such as Hangzhou Bay, the Yangtze River Estuary, the East China Sea, and the Bohai Sea in China. The database is organized in sliced format, consisting of image sections containing targets. Each slice is fixed at 1024x1024 pixels, with a sliding window of 400 pixels for slicing. The dataset comprises 3592 scene slices and 10127 SAR ship targets, covering 8 categories (the same as mentioned in subsection B), out of which 5422 targets have fine-grained category annotations. The dataset's annotation and division details are as follows.

*1) Annotation Attributes*

```
{
    "label": "Tanker", //Major category
    "label_s": "Oil_Chemical Tanker", //Subcategory
    "heading": 35, //Target orientation
    "difficult": 0, //Indicates difficult status
    "truncated": 0, //Indicates truncation status
    "shape_type": "polygon", //Box type
    "points": [ //Coordinates of the vertices
        [583, 500],
        [555, 474],
        [644, 381],
        [672, 407]
    ]
},
```

Fig. 16. Examples of ship target annotation attributes.

Ships are annotated with oriented bounding boxes, which are the smallest quadrilaterals fitting the ship areas, as illustrated in Fig. 16. Each target in the annotation file includes major and minor categories. The heading, derived from AIS information matching, is the angle from true north. 'Points' refer to the four vertices of the quadrilateral. In practice, the angle information of the ship can be obtained using different oriented box representations.

*2) Data Division*

The constructed dataset has also been divided into training and validation sets. Due to the sliding window strategy used during the slicing process, there is a possibility of repeated targets in slices from the same scene. Therefore, the dataset division is based on entire scenes, with training set slices from 20 scenes and the validation set from an independent set of 10 scenes. The training set includes 2555 slices and 7447 samples, of which 3230 samples have category annotations. The validation set comprises 1037 slices and 2680 samples, with 1475 samples having category annotations. The division also considers the diversity of sample categories. Fig. 17 shows the number of major categories in both the training and validation sets, covering all major categories.

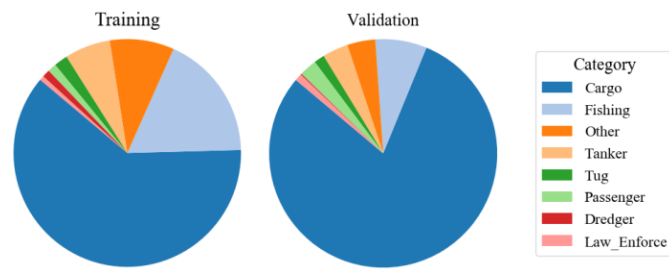

Fig. 17. Distribution of ship categories in training and validation sets.

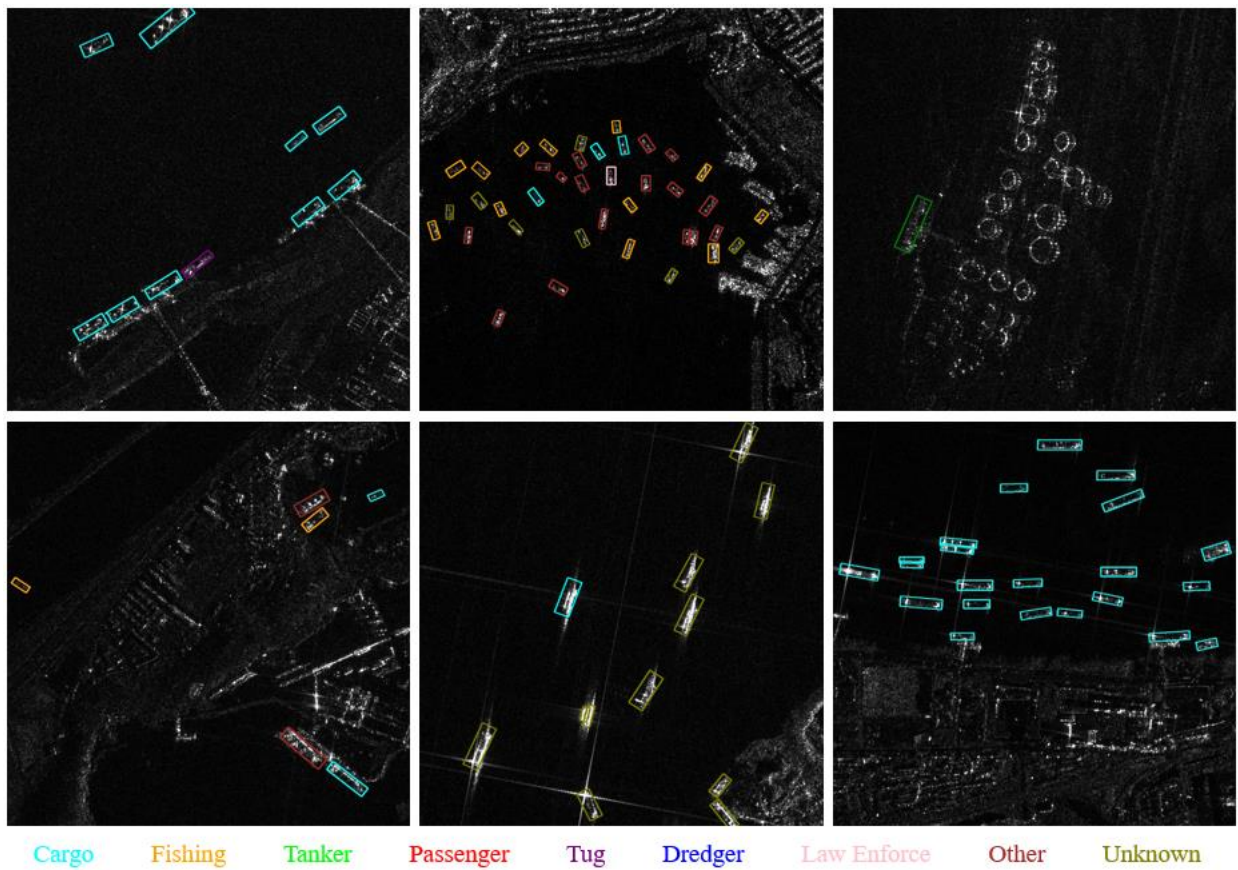

Fig. 18. Partial scene images and annotations in the dataset.

Fig. 18 shows some scene slices and annotated examples from the dataset, while Fig. 19 displays typical ship slices of different categories in the constructed dataset.

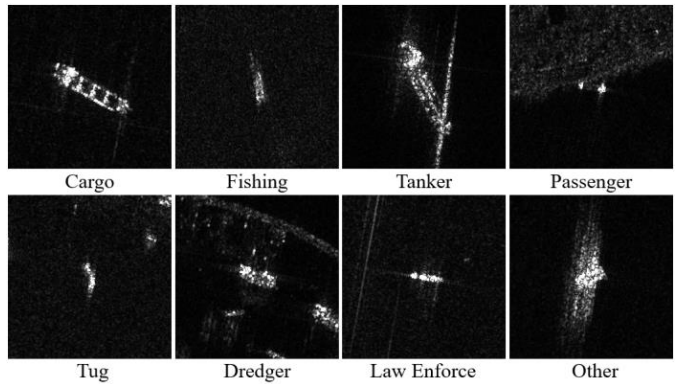

Fig. 19. Typical ship slices of each category in the dataset.

## VI. CONCLUSION

Addressing the issue of low automation in human-machine collaboration for target detection tasks in SAR images, a HitL-based learning framework is proposed in this paper for SAR ship detection. Specifically, within each iteration process of the framework, detection results are provided to users for correction, and the refined information is fed back to the model for iteration training, forming a continuous learning loop. Valuable sample selection and the interactive HMI are two core components of the loop. Firstly, active learning strategies tailored to SAR ship detection tasks are designed to maximize the value of samples recommended to users, significantly reducing the annotation workload. Secondly, for the HMI, efficient interactive operations and sample iteration attributes are developed based on the characteristics of SAR images, effectively channeling user operations back into the model. Furthermore, the proposed framework is applied to the construction of a SAR ship detection database, reducing the annotation costs and validating the effectiveness of the proposed method. Ultimately, a high-quality ship detection dataset of SAR images is established.